# Nanopore fabrication and characterization by helium ion microscopy

D. Emmrich[1], A. Beyer[1], A. Nadzeyka[2], S. Bauerdick[2], J. C. Meyer[3], J. Kotakoski[3], A. Gölzhäuser[1]

[1]*Physics of Supramolecular Systems and Surfaces, Bielefeld University, 33615 Bielefeld, Germany*

[2]*Raith GmbH, Konrad-Adenauer-Allee 8, 44263 Dortmund, Germany*

[3]*Faculty of Physics, University of Vienna, 1090 Vienna, Austria*

Abstract:

The Helium Ion Microscope (HIM) has the capability to image small features with a resolution down to 0.35 nm due to its highly focused gas field ionization source and its small beam-sample interaction volume. In this work, the focused helium ion beam of a HIM is utilized to create nanopores with diameters down to 1.3 nm. It will be demonstrated that nanopores can be milled into silicon nitride, carbon nanomembranes (CNMs) and graphene with well-defined aspect ratio. To image and characterize the produced nanopores, helium ion microscopy and high resolution scanning transmission electron microscopy were used. The analysis of the nanopores' growth behavior, allows inferring on the profile of the helium ion beam.

Nanopores in atomically thin membranes can be used for biomolecule analysis,[1] electrochemical storage,[2] as well as for the separation of gases and liquids.[3] All of these applications require a precise control of the size and shape of the nanopores. It was shown that the focused beam of a transmission electron microscope (TEM) is able to create nanopores in membranes of silicon nitride and graphene with diameters down to 2 nm.[4,5] Pores can be further shrunk in a TEM by areal electron impact.[6] However, the preparation of such nanopores in a TEM is time-consuming and is limited to small samples (~3 mm diameter) that fit into the microscope. Focused ion beam tools (FIB) offer more flexibility concerning the sample size and a higher milling speed. Among these FIB tools gallium liquid metal ion sources (LMIS) are widely used, allowing minimum sizes of 3 nm diameter for nanopores.[7] The development of a reliable gas field ionization source (GFIS) type allowed the construction of the helium ion microscope which surpasses the imaging and milling resolution of LMIS-based FIB tools.[8,9] First studies about milling with helium ions reported sample damage by amorphization and helium implantation during milling on bulk substrates.[10] The latter effect is absent on membranes, where nanopores with diameters of 2.6 nm were milled by HIM.[11] In all these reports, pores were created by single spot exposures. Here we present a different route to create small nanopores in membranes by milling circular patterns. Furthermore we are able to connect the growth of nanopores to the ion beam profile.

The work was performed with a Zeiss ORION Plus helium ion microscope, equipped with a Raith ELPHY Multibeam pattern generator. Pores were typically milled at beam energies of 35-40 keV at a working distance of 7-10 mm. Depending on the substrate and its cleanliness, the helium ion beam current was set to values of 0.5-6.2 pA. Secondary electrons (SE) as well as transmitted ions were used as signals for imaging. For the latter, a scanning transmission ion microscopy (STIM) extension was utilized.[12] For the high resolution imaging of nanopores, an ultra-high vacuum scanning transmission electron microscope (Nion UltraSTEM 100) provided atomic resolution images at a beam energy of 60 keV, while minimizing sample damage.[13] At the STEM the high-angle annular dark field signal (HAADF) was chosen. As membranes we used 30 nm thick silicon nitride membranes (Silson Ltd), single layer graphene sheets grown by CVD,[14] and 1 nm thick carbon nanomembranes (CNM). CNMs were made by electron irradiation of aromatic self-assembled monolayers leading to a two-dimensional cross-linking and the formation of mechanically free-standing membranes.[15-19]

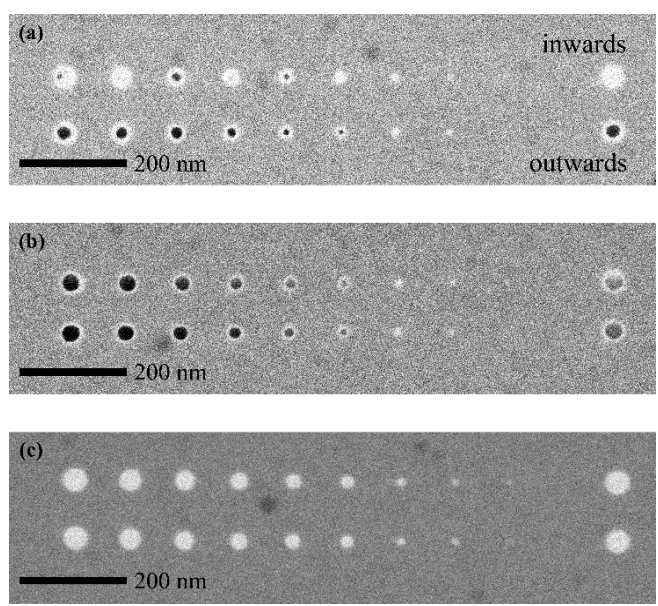

Figure 1: Three different nanopore arrays in silicon nitride imaged by HIM (transmission detection, STIM). (a) Writing a pattern of circular areas in concentric inward (top row) and concentric outward direction (bottom row) is accompanied by hydrocarbon deposition, depending on the writing path. (b) When repeating the pattern writing of (a) at another membrane position after cleaning the sample, no difference between inward and outward direction is observed. Note that (a) and (b) show fine details, but –unexpectedly- pores appear darker than the membrane. (c) Nanopores imaged with a cleaned, i.e. hydrocarbon free, STIM detector, showing an –expected- contrast with "bright" pores in a "dark" membrane.

Two approaches were made to mill nanopores: (i) keeping the beam at a constant position and unblanking it for a desired exposure time and (ii) exposing the sample with a predefined pattern. Fig. 1 shows nanopore arrays in silicon nitride which were created by patterns of circular areas. The arrays were then imaged with a HIM in the transmission (STIM) mode. At first sight, it is striking that although Fig. 1 (a) and (b) are bright field transmission images, the milled holes appear darker than the SiN membrane. This inversion of contrast is most likely due to hydrocarbons on the STIM detector. This detector generates a secondary electron signal from the transmitted helium ions. In presence of a high ion intensity, i.e. at openings in the SiN membrane, hydrocarbons on the detector can charge, which deflects the (low energy) secondary electrons. Surprisingly, this disturbance happens in a way that small features in the STIM images of nanopores, cf. Fig1a, b, can be better identified than with a clean STIM detector. After cleaning the sample holder in an argon plasma at 10 W for 30 min, the STIM detector operates "as expected" and the corresponding image shows "bright" open pores in a "dark" SiN membrane, cf. Fig, 1c.

Milling circular areas facilitate to obtain perfectly round nanopores of different sizes. Furthermore the membranes are thinner at the edges, creating pores with low aspect ratios. The milled pattern consists of circles with decreasing diameters from 40 nm (left) down to 2 nm (right) and a final 40 nm circle on the far right as a marker. The ion dose of $2.4 \times 10^{18}$ ions/cm in Figure 1 (a) and (b) is too low to mill the complete area, so that the 40 nm diameter in the exposure results in a 30 nm pore on the sample. Our aim is to thin the membrane until a small hole is breaking off right in the middle of the area. Figure 1a shows two lines of nanopores, patterned with the same ion dose but with different writing strategies. In the upper line, the beam moves in concentric circles inwards, in the lower line outwards. It can be seen that there is a clear difference between these writing directions. Writing inwards leads to closed or partially closed pores while writing in outward direction leads to more opened nanopores. The reason for this difference is the deposition of contaminant hydrocarbons on the pores during the milling process. In Fig. 1a the sample is introduced into the specimen chamber of the HIM after an air plasma treatment in the load lock for 8 min at 10 W, while in Fig. 1b the sample was cleaned for 23 min and stayed an additional 12 h in vacuum ($3 \times 10^{-7}$ mbar). While in Fig. 1b the same milling parameters were used as in Fig. 1a, the differences between the writing directions are not visible. Hence, the direction dependence must be related to the presence of hydrocarbons as well as to their surface diffusion on the sample. If the milling starts in the center of the pore, hydrocarbons are not able to diffuse into the center while the beam proceeds outwards. This keeps the pore center free of contamination and the milling rate is not reduced by hydrocarbon deposition. Conversely, if the writing starts at the outer boundary of a feature, hydrocarbons diffuse with the inwards moving beam.

On clean samples nanopores of 5 nm in diameter can be built by this writing strategy. When the ion dose is increased, the diameter of all pores increases and pores of diameters below 5 nm open at locations that had been closed before. However, during HIM imaging these nanopores further grow as the very thin material at their thin edge is sputtered away faster than material at the rim of larger high aspect ratio pores. We therefore expect that smaller pores can be detected by utilizing e.g. TEM below an electron energy of 120 keV which is known to be non-destructive on silicon nitride.[1]

A patterning strategy that employs the filling of geometrical objects thus provides reasonable results for a range of pore sizes and aspect ratios. However, for very thin membranes the aspect ratio is determined by the thickness of the membrane. In the case of graphene, taking control of the aspect ratio becomes meaningless, as the thickness is fixed. The following section thus deals with the creation of very small nanopores by spot exposures in carbon nanomembranes, graphene and silicon nitride. In this case the focused ion beam is kept at a fixed position and the pore size is controlled by the exposure time.

Figure 2 shows STEM micrographs of a nanopore array milled with the HIM into a CNM. The image shows a typical dose array used for determining the smallest pores which are opening at the *breakthrough dose*, just sufficient to form a hole. In case of Figure 2 the beam current was 1.4 pA and the shortest exposure time was 5 ms at the first spot in the 10x10 array in the upper right corner of (a). The exposure time increases from pore to pore by 5 ms from left to right in a single line, and by each line from top to bottom so that the highest exposure time is 500 ms in the lower right corner. For the STEM images the grey values of the HAADF signal are directly related to the thickness of the membrane. It has been shown that the HAADF signal is very sensitive to hydrocarbon contamination.[20] The CNM is an amorphous layer of cross-linked biphenyl molecules, and it also contributes to the thickness variation seen in the images. Light patches in (a) are assigned to residues of a polymer layer used in the transfer process.[16] Figure 2b shows a magnified area of the array with pores from two different lines. In horizontal direction the dose is increased in each line by the smallest step of 5 ms, corresponding to $4.7 \times 10^4$ ions/point while the dose difference between the lines is a factor of ten higher. The lower line shows what happens at doses high above the breakthrough dose. The upper line shows the behavior near the breakthrough dose: At the left position the membrane was not thinned while on the right position a pore was milled at a slightly increased dose. Even at lower dose on a different position (not shown here) it was possible to sputter through the membrane. The behavior at the breakthrough dose shows that slight differences in the membrane thickness determine whether or not a pore can be created. Figure 2c shows the smallest pore in this array at atomic resolution. The actual pore (see dashed circle) has a diameter of ~0.8 nm on the shorter axis and ~1.6 nm on the longer axis while its surrounding was thinned to atomic mono- or bilayers by the ion beam. The symmetry of the exposed area appears not circular. However, larger pores at longer exposure times indicate that the ion beam has a circular symmetry. Hence, it appears as the local stability of the molecular membrane influences the pore shape at very low doses.

Characteristic results for all investigated membranes are summarized in Table I. In all membranes nanopores below 4 nm could be created and detected. HIM and STEM were used for the analysis of the pore arrays. A comparison between HIM and STEM on both carbon based membranes shows that the smallest detectable pore size is around 3 nm in HIM, whereas STEM provides atomic resolution.

The breakthrough dose correlates with the thickness of the membranes and is the lowest for graphene. In contrary the silicon nitride membrane shows the highest sputter rate while graphene and CNM have lower ones. We attribute this to the higher thickness of the silicon nitride which increases the probability of a helium ion to hit a target atom. CNM and graphene show similar sputter rates, in agreement with their carbon based composition and their thinness.

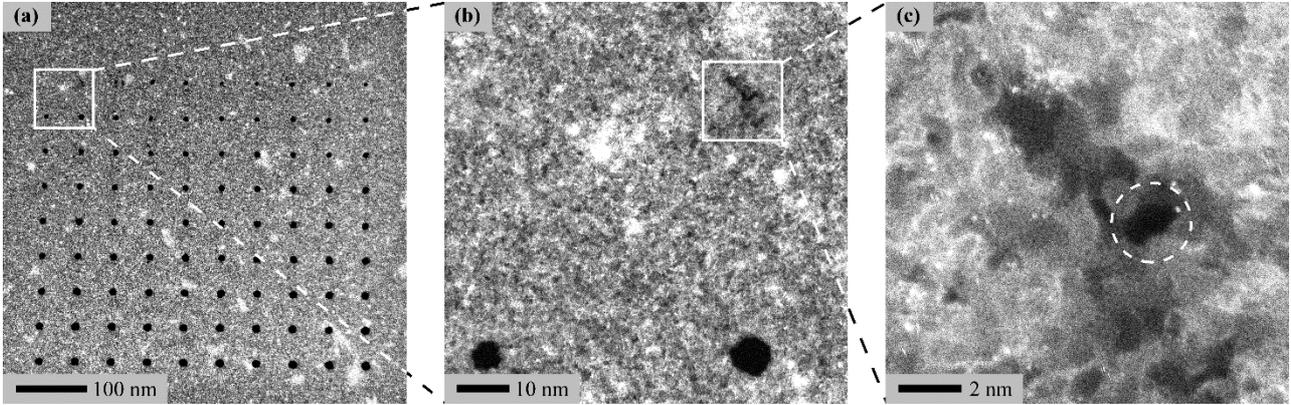

**Figure 2: STEM images (HAADF signal) of nanopores in a CNM. (a)** Overview of a spot exposure array. The dose increases from the top left to the bottom right corner. Bright patches result from residual polymer from the transfer. **(b)** The lower line pores received doses far beyond the breakthrough dose, while the upper line pores received doses at the breakthrough. At the upper left position no pore is visible, at the upper right position the dose ($5.6 \cdot 10^5$ ions/point) is just sufficient to create a defect in the membrane that is highlighted by a dashed circle in (c).

**Table I: Properties of the investigated membranes**

| Membrane | | Thickness | Microscope | Smallest diameter | Breakthrough dose | Sputter yield |
|---|---|---|---|---|---|---|
| | | nm | | nm | ions/point | |
| SiN | Non-conductive | 30 | HIM | 3.8 | $1.1 \times 10^6$ | 0.005 |
| CNM | Non-conductive | 1 | STEM | 1.3 | $4.5 \times 10^5$ | 0.001 |
| Graphene | Conductive | 0.34 | HIM | 2.9 | $3.7 \times 10^5$ | 0.001 |

In Figure 3 the pore diameters are plotted against the required ion doses. If the pores were not of circular shape, the diameter was averaged from a perfect circle with the same pore area. For graphene and silicon nitride membranes the curves were extracted from HIM images, STEM data were taken for CNMs. The curves show a steep increase of the pore size which gradually slows with increasing dose. The comparison of graphene and CNM shows the detection limits of the HIM, which obscures identification of pores below 3 nm, although both curves show a similar growth behavior. For silicon nitride we observe a slower pore size growth in agreement with the calculated sputter yield.

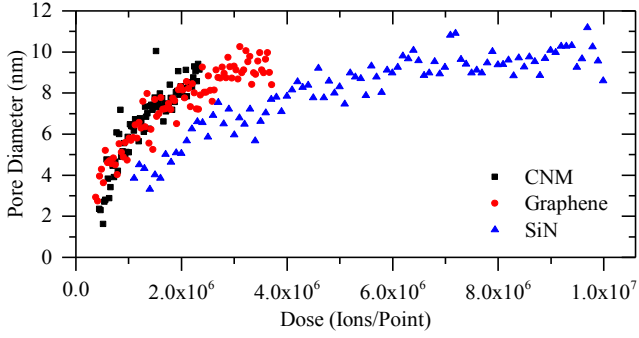

**Figure 3:** Diameters of nanopores milled with helium ions into different membranes extracted from microscope images. In the case of graphene and silicon nitride the sizes were determined on HIM images, for CNM STEM data was used.

The growth behavior of nanopores is further related to the profile of the milling ion beam, i.e. its point spread function (PSF). In resist based lithography with light and ions, it has been shown that the PSF can be deduced from spot exposure arrays.[21,22] We can adopt this concept for nanopore milling: When the breakthrough dose is exceeded at a certain point in the PSF, a hole is created. The diameter of a nanopore is therefore related to the width of the PSF at a certain height. Varying the exposure time changes the amplitude of the PSF:[21]

$$PSF(r_n) \times t_n = T_H \quad \Rightarrow \quad PSF(r_n) = T_H/t_n$$

Here $r_n$ is the radius of a pore, $t_n$ the exposure time for this pore and $T_H$ the breakthrough dose. The PSF can be calculated by plotting the reciprocal exposure time (which works as a normalization factor) against the pore radius. In our case we did the normalization not on the exposure time but on the ion dose to compensate different beam currents on different samples as shown in Figure 4. Since the membranes are ultimately thin, this measures the pristine ion beam profile without any contribution from secondary electrons or backscattered ions that may broaden the beam.

Figure 4 (a) shows beam profiles for CNM and silicon nitride, derived from the data in Figure 3. In a first approximation a Gaussian distribution is fitted to the data, having its maximum at 0 nm. The beam currents in (a) are comparable for both data sets, meaning the beam profile is expected to be the similar. In both cases the full width at half maximum (FWHM) gave identical results to the second decimal place. We thus could deduce identical beam profiles from milling membranes of different composition, sputter yield and thickness. We also found the dose of the HIM imaging to be sufficiently low to not widen the pores, as the STEM analysis for CNMs and the HIM analysis for SiN show similar beam profiles.

Figure 4 (b) compares beam profiles derived from arrays in graphene written at different beam currents. To vary the beam current, adjustments on the condenser lens of the HIM were done while we still use the 10 μm aperture. One can see that the lower beam current results in a much steeper slope having a FWHM of 3.82 nm, while the higher beam current has a FWHM of 7.90 nm.

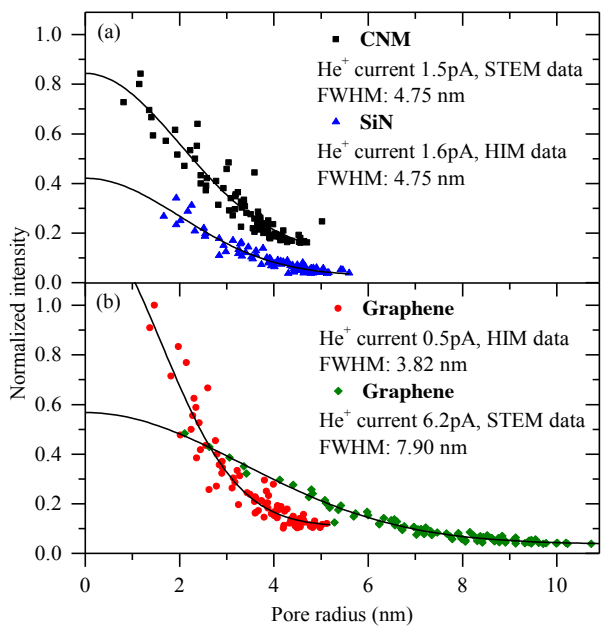

**Figure 4: Beam profiles of the helium ion beam determined from nanopore arrays on different substrates. As a first approximation a Gaussian distribution is assumed. The same beam setup at the HIM (a) leads to identical results for the FWHM of 4.75 nm, even on different membranes. An increase of the beam current is at the expense of the beam width (b).**

In summary, we demonstrated two approaches of creating nanopores into membranes of different material. With milling geometrical patterns we can vary the thickness of the pore's rim in silicon nitride, while realizing pore diameters down to 5 nm. With spot exposures we are able to mill smaller features and achieve pore diameters down to 1.3 nm. We evaluated the sputter yield for our membranes and deduced the beam profile from the growth of the nanopores. Remarkably, the profile can be also determined if we use the HIM for both milling and imaging. Milling well defined nanopores in the demonstrated size range as well as the improved knowledge of the helium ion beam characteristic will facilitate the application of such small pores in other scientific research areas like the analysis of biomolecules with nanopore-based approaches.

**Acknowledgements:**

This work was financially supported by the German Bundesministerium für Wirtschaft. This work was further conducted within the framework of the COST Action CM1301 (CELINA). The research leading to these results has also received funding from the European Union Seventh Framework Programme under grant agreement n°604391 Graphene Flagship. We acknowledge Wiener Wissenschafts-, Forschungs- und Technologiefonds (WWTF) project MA14-009 and European Research Council (ERC) Project No. 336453-PICOMAT.